\documentclass[3p,twocolumn]{elsarticle}

\journal{J. Crystal Growth}

\usepackage{graphicx}
\usepackage{amssymb}
\usepackage{amsmath}

\begin{document}



\begin{frontmatter}

\title{The melting behavior of lutetium aluminum perovskite LuAlO$_3$}

\author{Detlef Klimm}\ead{klimm@ikz-berlin.de}

\address{Leibniz Institute for Crystal Growth, Max-Born-Str. 2, 12489 Berlin, Germany}

\begin{abstract}
DTA measurements with mixtures of aluminum oxide and lutetium oxide around the 1:1 perovskite composition were performed up to $1970^{\,\circ}$C. A peak with onset $1901^{\,\circ}$C was due to the melting of the eutectic Lu$_4$Al$_2$O$_9$ (monoclinic phase) and LuAlO$_3$ (perovskite). Neither peritectic melting of the perovskite nor its decomposition in the solid phase could be resolved experimentally. The maximum of the eutectic peak size is near $x=0.44$, on the Lu-rich side of the perovskite, which is consistent with the conclusion that LuAlO$_3$ melts peritectically at ca. $1907^{\,\circ}$C as proposed by Wu, Pelton, J. Alloys Compd. 179 (1992) 259. Thermodynamic equilibrium calculations reveal, that under strongly reducing conditions (oxygen partial pressure $<10^{-13}$\,bar) aluminum(III) oxide can be reduced to suboxides or even Al metal. It is shown that under such conditions a new phase field with liquid Al can appear.
\end{abstract}


\end{frontmatter}


\section{Introduction}
\label{sec:Introduction}

The pseudo binary phase diagrams Al$_2$O$_3$--RE$_2$O$_3$ (RE stands for a rare earth element from La to Lu, or Y) contain up to four intermediate compounds REAl$_{11}$O$_{18}$ ($\beta$-alumina type, stable only for the larger RE$^{3+}$ from La to Eu), RE$_3$Al$_5$O$_{12}$ (garnet type, stable only for the smaller RE$^{3+}$ starting with Eu \cite{Garskaite07}), REAlO$_3$ (orthorhombic distorted perovskite type), and the monoclinic ($P\,2_1/c$) RE$_4$Al$_2$O$_9$ that were recently shown to exist for all RE$^{3+}$ \cite{Dohrup96}. Bulk single crystals from many of these compounds can be grown by conventional techniques such as Czochralski or Bridgman, and find applications e.g. in laser technology or as scintillators. The versatility of such crystals is enhanced by the fact that all of them can easily be doped with other RE$'_2$O$_3$ if the radii of RE$^{3+}$ and RE$^{'3+}$ are not too different.

Kaminskii et al. \cite{Kaminskii76} have grown Nd$^{3+}$:LuAlO$_3$ ($\approx1$\% doping) single crystals using the Czochralski technique (Lu$_2$O$_3$:Al$_2$O$_3$ = 1:1 starting material, Ar or N$_2$ atmosphere, Ir crucible, pulling rate $2-5$\,mm/h, rotation $20-45$\,rpm, optimum growth direction $\left[112\right]$). The resulting single crystals of several millimeter diameter and several centimeter length were mainly used for a thorough spectroscopic characterization. The space symmetry group was found to be $D^{16}_{2h}=P\,bnm$ with $Z=4$ and $a_0=5.100(3)$\,\AA, $b_0=5.324(2)$\,\AA, $c_0=7.294(1)$\,\AA. Occasionally, inclusions of Lu$_3$Al$_5$O$_{12}$ or Lu$_2$O$_3$ were found. Petrosyan et al. \cite{Petrosyan98} reported the Bridgman growth of Ce$^{3+}$:LuAlO$_3$ ($\leq1$\% doping) single crystals for scintillator applications (Mo crucible, Ar atmosphere with $\leq30$\% H$_2$, pulling rate $0.5-5$\,mm/h, diameter $\leq12$\,mm, length $\leq70$\,mm). Sometimes Mo inclusions ($1-6$\,$\mu$m sized platelets) were observed. Other unidentified inclusions of smaller size ($0.1-1$\,$\mu$m) in the last grown sections of heavily doped crystals were assumed to be a result of constitutional supercooling. Occasionally, gas bubble inclusions were found along the crystals' central axis.

\begin{figure}[htb]
\includegraphics[width=0.46\textwidth]{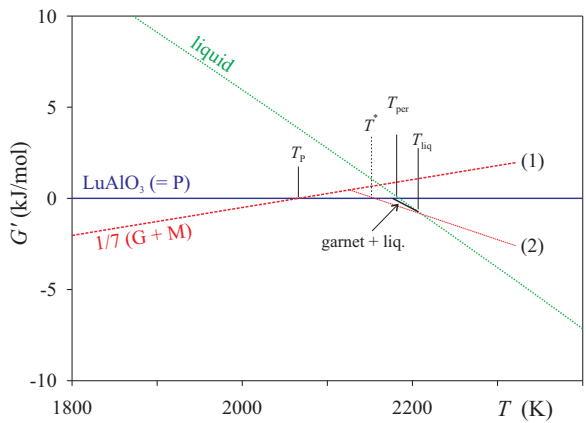}
\caption{Gibbs free enthalpy $G'$ for different phases and phase mixtures with identical composition LuAlO$_3$, compared with the reference state perovskite (P). 1/7~(G+M) with label (1) is the product side of (\ref{eq:dis}), label (2) would be the hypothetical case that perovskite is formed for $T>T_\mathrm{P}$ from garnet (G) + monoclinic phase (M), and decomposes for $T>T^*$ again to G+M. The dotted green line is pure liquid, and the short full line between $T_\mathrm{per}$ and $T_\mathrm{liq}$ is the mixture G+liq. between peritectic decomposition and liquidus. Data from \cite{FactSage5_5}.}
\label{fig:G_vs_T}
\end{figure}

It was reported that LuAlO$_3$ decomposes upon heating to the garnet Lu$_3$Al$_5$O$_{12}$ and the monoclinic Lu$_4$Al$_2$O$_9$ or even Lu$_2$O$_3$ \cite{Szupryczynski05,Ding07}. Recently Petrosyan et al. \cite{Petrosyan06} explained this observation by the assumption that LuAlO$_3$ is stable only in a limited temperature range $1750^{\,\circ}$C $\leq T \leq 1930^{\,\circ}$C and decomposes for higher and lower $T$ to Lu$_4$Al$_2$O$_9$ and Lu$_3$Al$_5$O$_{13}$. The decomposition of one phase A to two other phases B and C identically below and above some finite stability range of A is not strictly forbidden. However such decomposition is not very likely for thermodynamic reasons, if all phases A, B, C possess fixed stoichiometry: It would require unusually sharp bends in the $G(T)$ functions of the corresponding phases. Such behavior corresponds with the hypothetical curve (2) in Fig.~\ref{fig:G_vs_T}, whereas all other curves in this $G'(T)=G(T)-G_\mathrm{P}(T)$ diagram ($G_\mathrm{P}(T)$ = perovskite data) were calculated with FactSage \cite{FactSage5_5} from tabulated thermodynamic values. In thermodynamic equilibrium the phase(s) with lowest $G(T)$, for a given composition $x$, are stable, and from Fig.~\ref{fig:G_vs_T} it is obvious that (G+M) transforms to (P) at $T_\mathrm{P}$, further to (G+liq.) at $T_\mathrm{per}$ (peritectic melting), and finally to liquid at $T_\mathrm{liq}$. Almost straight lines for $G(T)=H-T\,S$ are realistic for each single phase or phase mixture, as opposed to a sharp bend, which is indicated with curve (2), as enthalpy $H$ and entropy $S$ usually do not vary much with $T$. If one phase (as is indicated for the garnet by a dashed line in \cite{Petrosyan06}) possesses some finite homogeneity range, the claims given above are not so strictly valid, but nevertheless strongly bent $G(T)$ curves seldom occur.

In \cite{Petrosyan06} only the garnet melts congruently at $2060^{\,\circ}$C, and the monoclinic phase melts peritectically at $2000^{\,\circ}$C under the formation of Lu$_2$O$_3$. These claims were summarized in a phase diagram Lu$_2$O$_3$--Al$_2$O$_3$ that differs considerably from the thermodynamic assessment by Wu and Pelton \cite{Wu92} where LuAlO$_3$ melts incongruently at $1907^{\,\circ}$C under the formation of Lu$_3$Al$_5$O$_{12}$. Later Kanke and Navrotsky \cite{Kanke98} reported enthalpy measurements by drop-in calorimetry with different RE-Al oxides, but LuAlO$_3$ was not measured in this report. It was claimed instead that LuAlO$_3$ could only be prepared under high pressure (which is obviously not true \cite{Kaminskii76,Petrosyan98}) --- as the stability was said to be limited by the disproportionation reaction
\begin{equation}
\mathrm{LuAlO}_3 \rightleftarrows \frac{1}{7} \mathrm{Lu}_3\mathrm{Al}_5\mathrm{O}_{12} + \frac{1}{7} \mathrm{Lu}_4\mathrm{Al}_2\mathrm{O}_9 \label{eq:dis}
\end{equation}
which would be in agreement with \cite{Petrosyan06}. Unfortunately, equilibria with the monoclinic phase were not discussed further in \cite{Kanke98} and instead the decomposition of perovskite to garnet and Lu$_2$O$_3$ was discussed quantitatively (Fig.~7 in \cite{Kanke98}). The dissociation of several rare earth aluminum perovskites REAlO$_3$ (RE = Gd, Ho, Er, Y, Tm, Yb, Lu) according reaction (\ref{eq:dis}) was found by Bondar' et al. \cite{Bondar79} who performed X-ray phase analysis of annealed poly- and single crystals. The decomposition rate was reported to depend on temperature and annealing atmosphere, and, especially for RE = Tm, Yb, and Lu and at very high $T\geq1830^{\,\circ}$C, a reducing atmosphere accelerates the decomposition.

The present paper reports differential thermal analysis (DTA) measurements with compositions around LuAlO$_3$ that were performed to clarify the contradictions mentioned above.


\section{Experimental and Results}
\label{sec:Experimental}

DTA measurements were performed with a NETZSCH STA 409C (graphite furnace, DTA sample holder with thermocouples W/Re). Lu$_2$O$_3$ and Al$_2$O$_3$ powders ($\geq99.99$\% purity) were mixed in a molar ratio 1:1 (molar fraction of Al$_2$O$_3$ $x=0.500$) in a mortar and $\approx20$\,mg of the mixtures were filled in DTA crucibles made of tungsten. The measurements were performed in flowing argon (99.999\% purity, 40\,ml/min) with heating/cooling rates of $\pm10$\,K/min up to $1970^{\,\circ}$C. Homogenization of the samples was obtained during a first DTA heating/cooling cycle, and only the subsequent DTA runs are used in the following for the derivation of phase equilibria. This procedure has the benefit that the chemical composition of the sample is always known exactly -- which is in contrast to ex situ preparation by melting and crystallization, that can lead to segregation of the chemical components. Unfortunately, the small sample mass, some of which sticks tightly to the crucible wall, does not allow subsequent phase analysis by X-ray diffraction techniques. It should be noted that the DTA apparatus that was used here is virtually the same as was used by Ding et al. \cite{Ding07}. The minor differences are only with respect to some electronic parts; but the furnace, sample holder with thermocouples, crucible, heating rate, and atmosphere are identical.

\begin{figure}[htb]
\includegraphics[width=0.46\textwidth]{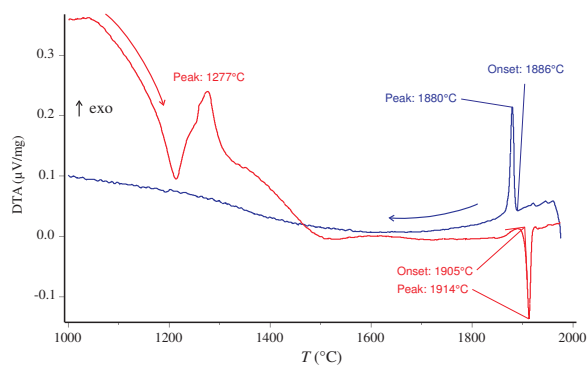}
\caption{DTA second heating and cooling curves for $x=0.500$ (= LuAlO$_3$).}
\label{fig:heat-cool}
\end{figure}

Fig.~\ref{fig:heat-cool} shows the second heating and cooling curves for the perovskite composition. Eleven other samples in a concentration range $0.352\leq x\leq0.615$ were prepared by adding minor quantities of Lu$_2$O$_3$ or Al$_2$O$_3$, respectively, to the 1:1 mixture. The main features of Fig.~\ref{fig:heat-cool} are almost identical with the DTA heating and cooling curves of crystalline LuAlO$_3$:Ce that are shown in Fig.~2 of Ding's paper \cite{Ding07}. It should be noted that there the ``exo'' direction was chosen downwards, whereas throughout this paper ``exo'' is upwards, in agreement with the ASTM E~472-86 standard. Another issue of \cite{Ding07}, however, is critical: DTA peak temperatures were taken there to construct the phase diagram, but this is principally incorrect. The start of melting corresponds to extrapolated onset temperatures, and peak temperatures lead to a systematic overestimation of melting points $T_\mathrm{f}$ during heating, and to underestimation of $T_\mathrm{f}$ if cooling curves are used \cite{Hohne96}.

The broad effects in the heating curve for $T$ $\approx1200-1300^{\,\circ}$C were attributed by Ding et al. \cite{Ding07} to the partial decomposition of the perovskite, as described in equation (\ref{eq:dis}), and this interpretation was apparently justified by high temperature X-ray diffraction patterns. Indeed, this explanation seems possible, but a phase transformation of the monoclinic phase Lu$_4$Al$_2$O$_9$ is an alternative explanation: For the isomorphous yttrium compound Y$_4$Al$_2$O$_9$ Yamane et al \cite{Yamane98} found by neutron diffraction a phase transformation from a low-$T$ $P2_1/c$ phase to a high-$T$ phase with identical space symmetry group. This martensitic transformation is diffusionless and takes place by a shear mechanism near $1370^{\,\circ}$C. The transformation rate depends on the mechanical stress state of the crystallites, and consequently on the grain size, as is typical for martensitic transformations. Unfortunately, no data are published so far for Lu$_4$Al$_2$O$_9$, but for all other small rare earth elements (starting from RE = Sm) the monoclinic RE$_4$Al$_2$O$_9$ undergo a similar transformation between $1044^{\,\circ}$C (RE = Sm) and $1300^{\,\circ}$C (RE = Yb) \cite{Yamane95}. All Lu$_2$O$_3$--Al$_2$O$_3$ phase diagrams published so far \cite{Ding07,Petrosyan06,Wu92} assume that LuAlO$_3$ decomposes slowly below $\approx1770^{\,\circ}$C, and consequently traces of Lu$_4$Al$_2$O$_9$ within the solid could be responsible for the effects between 1200 and $1300^{\,\circ}$C.

The large peaks in the heating and cooling curves near $1900^{\,\circ}$C result from melting or crystallization of the sample. The onset temperature of the crystallization peak in Fig.~\ref{fig:heat-cool} is lower than the onset of the melting peak by the supercooling $\Delta T=19$\,K, and for other compositions $\Delta T$ is even larger. The reproducibility and comparability was better for DTA heating curves, compared with cooling curves. Only second heating curves, that were measured after the first homogenization run, are used in the following for the investigation of phase equilibria.

\begin{figure}[htb]
\includegraphics[width=0.46\textwidth]{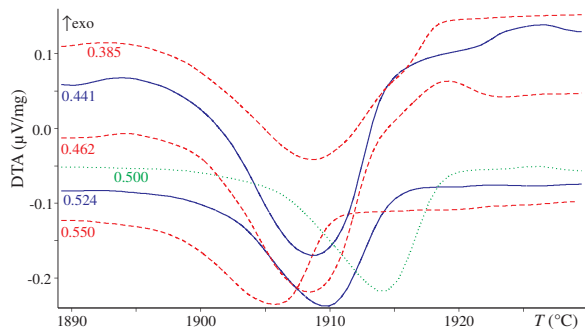}
\caption{DTA heating curves for 6 compositions around $x=0.500$ (= LuAlO$_3$, composition $x$ given as parameter).}
\label{fig:dta}
\end{figure}

The second heating DTA curves that were obtained with some samples around the LuAlO$_3$ composition ($x=0.500$) are shown in Fig.~\ref{fig:dta}. It turns out that all samples showed one melting peak with extrapolated onset $T_\mathrm{on}=1901\pm3^{\,\circ}$C. For $x=0.500$, $T_\mathrm{on}=1904^{\,\circ}$C was measured, and this value is not significantly larger in comparison with the other compositions $0.352\leq x\leq0.550$. (The latter is the last composition where this peak could be observed.) The peak area has a maximum value $A=14.4$\,$\mu$Vs/mg for $x=0.441$ and becomes smaller to both sides: $A=8.45$\,$\mu$Vs/mg for $x=0.352$, $A=8.7$\,$\mu$Vs/mg for $x=0.500$, $A=5.7$\,$\mu$Vs/mg for $x=0.550$. For some samples, the peak had a small shoulder on the high-$T$ side that could indicate the spacing between eutectic melting and liquidus temperatures (e.g. $x=0.441$ in Fig.~\ref{fig:dta}). No additional peaks could be found for any sample up to $1970^{\,\circ}$C. A second peak due to the peritectic decomposition of LuAlO$_3$ should be expected for Al-rich compositions, but the thermal difference of the eutectic and the peritectic is only $\approx8-10$\,K (see Fig.~\ref{fig:pd}) and could not be resolved due to the limited resolution at such high $T$.

It would be desirable to perform DTA measurements in the whole system from Lu$_2$O$_3$ to Al$_2$O$_3$, and especially around $x=\frac{1}{3}$ (Lu$_4$Al$_2$O$_9$) and $x=0.625$ (Lu$_3$Al$_5$O$_{12}$), but unfortunately the melting points for all 4 compounds are $>2000^{\,\circ}$C and cannot be reached with the DTA apparatus that was available.


\section{Discussion}
\label{sec:Discussion}

Both Petrosyan et al. \cite{Petrosyan06} and Wu et al. \cite{Wu92} report that LuAlO$_3$ is an intermediate phase between Lu$_4$Al$_2$O$_9$ and Lu$_3$Al$_5$O$_{12}$. The phase diagram that is reported by the former authors (Fig.~1 in \cite{Petrosyan06}) shows an eutectic point $x_\mathrm{eut}\approx0.5$, $T_\mathrm{eut}\approx1960^{\,\circ}$C between Lu$_4$Al$_2$O$_9$ (incongruently melting at $2000^{\,\circ}$C) and Lu$_3$Al$_5$O$_{12}$ (congruently melting at $2060^{\,\circ}$C). In contrast, the phase diagram by Wu et al. (Fig.~17 in \cite{Wu92}), based on a thermodynamic assessment, shows LuAlO$_3$ melting incongruently at $1907^{\,\circ}$C under formation of Lu$_3$Al$_5$O$_{12}$ (congruently melting at $2043^{\,\circ}$C). Between Lu$_4$Al$_2$O$_9$ (congruently melting at $2040^{\,\circ}$C) and LuAlO$_3$ a eutectic point ($x_\mathrm{eut}=0.46$, $T_\mathrm{eut}=1897^{\,\circ}$C) is shown.

The current DTA measurements showed no thermal effects near $1960^{\,\circ}$C, but a strong melting peak near $1901^{\,\circ}$C instead. The maximum peak size was found near $x=0.441$. Both $T$ and $x$ correspond well with the eutectic point that was reported by Wu et al. \cite{Wu92}. The current results are not in agreement with the results of Petrosyan et al. \cite{Petrosyan06} where the eutectic is proposed at higher $T$ and at $x=0.50$. The disproportionation reaction (\ref{eq:dis}) cannot explain the DTA peaks in Fig.~\ref{fig:dta} for the following reasons:
\begin{enumerate}
	\item The measured DTA peaks are by $\approx20$\,K too low.
	\item The maximum peak size was measured slightly to the Lu-rich side ($x\approx0.44$) from the LuAlO$_3$ composition where it should be if the perovskite decomposed in the solid phase.
	\item Such a strong thermal effect with large consumption of heat is expected to be the result of a melting process rather than a process between solid phases only. Indeed it could be seen that the DTA samples were really molten directly after passing the peak, if the DTA measurement was stopped there.
\end{enumerate}

It can be concluded that under the current experimental conditions the phase diagram of Wu et al. \cite{Wu92} is correct. However, the question should be discussed why different results were found by others: Petrosyan et al. \cite{Petrosyan06} write that their measurements were performed under argon/hydrogen atmosphere, with unspecified composition. Moreover, it is claimed that ``\ldots Lu and Al have stable oxidation states (III) \ldots (and) changes in phase states of condensed systems will not depend on the change \ldots of the partial pressure of oxygen in the co-existing gaseous phase''.

In a previous paper \cite{Petrosyan81} Petrosyan et al. used a mixture of 20\,Vol.\% H$_2$ with 80\,Vol\% Ar and one can assume similar conditions here. If such gas mixture equilibrates with LuAlO$_3$ a resulting $p_{\mathrm{O}_2}(T)=1.1\times10^{-13}$\,bar at $1900^{\,\circ}$C or $p_{\mathrm{O}_2}(T)=1.7\times10^{-12}$\,bar at $2000^{\,\circ}$C, respectively, can be calculated \cite{FactSage5_5}. If 30\,Vol.\% H$_2$ in Ar is used instead \cite{Petrosyan98}, these $p_{\mathrm{O}_2}(T)$ are further scaled down by a factor $\approx0.75$. It should be noted that under such experimental conditions O$_2$ is mostly dissociated: $p_\mathrm{O}(T)=7.7\times10^{-10}$\,bar at $1900^{\,\circ}$C and $p_\mathrm{O}(T)=5.6\times10^{-9}$\,bar at $2000^{\,\circ}$C. For the growth of doped or undoped sapphire crystals ($\alpha$-Al$_2$O$_3$, $T_\mathrm{f}=2054^{\,\circ}$C), the formation of bubbles is a well known issue \cite{Saito86,Song05b}. Aluminum suboxides Al$_2$O(gas), AlO(gas), and Al(gas), which are formed especially under high $T$ and low $p_{\mathrm{O}_2}$ are involved in the formation of such gaseous inclusions \cite{Uecker06,Zhang08b}. It is interesting to note that gas bubble inclusions have been seen in Ce:LuAlO$_3$ crystals under Petrosyan's growth conditions \cite{Petrosyan98} with up to 30\,Vol.\% H$_2$ in Ar.

\begin{figure}[htb]
\includegraphics[width=0.46\textwidth]{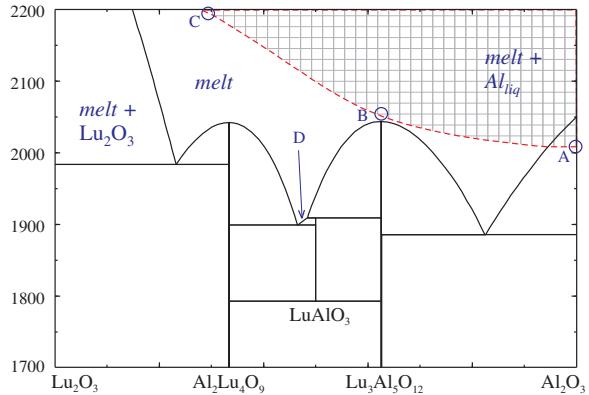}
\caption{The phase diagram Lu$_2$O$_3$--Al$_2$O$_3$, calculated with FactSage and Wu's data \cite{FactSage5_5,Wu92} for a oxygen partial pressure $p_{\mathrm{O}_2}>10^{-13}$\,bar (solid lines). For a lower $p_{\mathrm{O}_2}=2.5\times10^{-14}$\,bar a new phase field with liquid Al appears for high $T$ and high Al concentrations. Vapor pressures for several species at points A -- D are given in Table~\ref{tab:pressure}.}
\label{fig:pd}
\end{figure}

Fig.~\ref{fig:pd} shows with solid lines the phase diagram Lu$_2$O$_3$--Al$_2$O$_3$ as reported by Wu and Pelton \cite{Wu92,Acers04}, where LuAlO$_3$ melts incongruently at $1907^{\,\circ}$C. It turns out that this diagram is valid only for sufficiently high $p_{\mathrm{O}_2}\gtrsim10^{-13}$\,bar. For the Ar/H$_2$ mixtures mentioned above, however, the calculated oxygen partial pressure is very close to this critical limit. If $p_{\mathrm{O}_2}$ is slightly lower, Al$_2$O$_3$-rich melts are reduced and Al(liq) appears as an additional phase in the top right corner of Fig.~\ref{fig:pd}. The new boundary (dashed line) separating the phase field ``melt + Al$_\mathrm{liq}$'' from ``melt'' moves to lower $T$ if $p_{\mathrm{O}_2}$ decreases.

Already for $p_{\mathrm{O}_2}=2.5\times10^{-14}$\,bar, as shown in Fig.~\ref{fig:pd}, the ``melt + Al$_\mathrm{liq}$'' field touches the Al$_2$O$_3$ liquidus line. This means that the liquidus, starting from the eutectic point Lu$_3$Al$_5$O$_{12}$/Al$_2$O$_3$, does not reach the melting point of pure Al$_2$O$_3$ ($x=1.00, T_\mathrm{f}=2054^{\,\circ}$C). Instead, it bends horizontally at $x=0.95, T=2007^{\,\circ}$C in this case. During the calculations that lead to Fig.~\ref{fig:pd} the gas phase was taken into account as an ideal mixture. Thus it was possible to calculate the vapor pressure of some relevant species at several points, and the results are shown in Table~\ref{tab:pressure}. Al$_2$O and Al are the most important species for aluminum, and LuO and Lu for lutetium. Along the dashed phase boundary the vapor pressure of lutetium bearing species is $\lesssim10^{-7}$\,bar at points A and B (close to phase fields with solid phases), and reaches not more than $p_\mathrm{V}=4.3\times10^{-5}$~bar (Lu$_\mathrm{gas}$) even for the highest $2200^{\,\circ}$C shown here (point C). For such low $p_\mathrm{V}$ the evaporation of lutetium is expected to be very small.

The $p_\mathrm{V}$ for Al$_2$O and Al, contrarily, are much larger at every point and reach values as high as $\approx200$\,mbar at points A--C. Even at point D, in the middle of the LuAlO$_3$ liquidus, the combined vapor pressure of the aluminum bearing species
\begin{equation}
	p_\mathrm{V}^{\mathrm{total,Al}} = 2 p_\mathrm{V}^{\mathrm{Al}_2\mathrm{O}} + p_\mathrm{V}^{\mathrm{Al}}
\end{equation}
reaches 0.9\,mbar. This is already sufficiently high, and considerable evaporation of Al cannot be ruled out -- especially if the melt is overheated.

\begin{table}[htb]
\caption{Vapor pressures ($p_\mathrm{V}$ in bar, $p_{\mathrm{O}_2}=2.5\times10^{-14}$\,bar) for several gaseous species at points A -- D shown in Fig.~\ref{fig:pd}.}
\footnotesize
\begin{tabular}{ccccc}
\hline
  & Al$_2$O            & Al                 &      LuO           & Lu \\
\hline
A & $2.2\times10^{-1}$ & $5.7\times10^{-2}$ & $2.0\times10^{-9}$ & $1.2\times10^{-9}$ \\
B & $1.8\times10^{-1}$ & $7.8\times10^{-2}$ & $1.1\times10^{-7}$ & $1.0\times10^{-7}$ \\
C & $9.8\times10^{-2}$ & $1.9\times10^{-1}$ & $1.1\times10^{-5}$ & $4.3\times10^{-5}$ \\
D & $1.6\times10^{-4}$ & $5.7\times10^{-4}$ & $1.7\times10^{-8}$ & $3.3\times10^{-9}$ \\
\hline
\end{tabular}
\normalsize
\label{tab:pressure}
\end{table}


\section{Conclusions}
\label{sec:Conclusions}

Under sufficiently high oxygen partial pressure $p_{\mathrm{O}_2}>10^{-13}$\,bar the melting behavior of LuAlO$_3$ can be described by the Lu$_2$O$_3$--Al$_2$O$_3$ phase diagram of Wu and Pelton \cite{Wu92,Acers04} (Fig.~\ref{fig:pd}). Under strongly reducing conditions, however, Al$_2$O$_3$ is reduced partially and aluminum bearing species reach high volatility. This can lead to the formation of a new phase field with Al(liq) in the phase digram as well as to the enhanced evaporation of Al from the melt, resulting in a concentration shift. The liquidus of LuAlO$_3$ in Fig.~\ref{fig:pd} extends only from $x=0.50$ to $x_\mathrm{eut}=0.46$, resulting in a tiny crystallization window of only 4\,mol\%. If the very high heating rates of Petrosyan et al. \cite{Petrosyan98,Petrosyan06,Petrosyan81} up to 3000\,K/min are taken into account, some degree of overheating seems to be realistic. This, together with strong gas convection, may be responsible for aluminum loss and for the claim of the perovskite decomposition in the solid phase following (\ref{eq:dis}), or even under the formation of Lu$_2$O$_3$.

The main difference of the Lu$_2$O$_3$--Al$_2$O$_3$ phase diagrams that was published by Petrosyan et al. \cite{Petrosyan06}, compared with the diagram by Wu, Pelton \cite{Wu92} (that can also be found in \cite{Ding07,Acers04}) is the decomposition of solid LuAlO$_3$ perovskite to a phase mixture of solid Lu$_4$Al$_2$O$_9$ (monoclinic phase) and solid Lu$_3$Al$_5$O$_{12}$ below and above some critical temperatures. The decomposition seems unquestionable for small $T\lesssim1800^{\,\circ}$C, but no indications were found here for the equilibrium decomposition of the perovskite to garnet and monoclinic phase at $T\gtrsim1930^{\,\circ}$C. Instead, peritectic melting at $\approx1907^{\,\circ}$C, as already reported by Wu, Pelton \cite{Wu92}, was confirmed. Under strongly reducing atmosphere, as is sometimes used for the growth of LuAlO$_3$ crystals, Al$^{3+}$ can be partially reduced to metallic aluminum, or to aluminum suboxide. Only in this case, the evaporation of Al or Al$_2$O may lead to the decomposition of LuAlO$_3$ already below its peritectic melting point.


\section*{Acknowledgements}

The author is indebted to T. \L ukasiewicz and M. \'{S}wirkowicz (ITME Warszawa) for stimulation to this work and to D.\,A. Pawlak (ITME Warszawa) and S. Ganschow (IKZ Berlin) for discussion and reading the manuscript. This work was supported by the EU Commission in the Seventh Framework Programme through the ENSEMBLE project (Grant Agreement Number NMP4-SL-2008-213669).


\end{document}